\documentclass[10pt,a4paper]{article}
\usepackage{fontenc}
\usepackage{xcolor}
\usepackage{amsmath}
\usepackage{amsfonts}
\usepackage{amssymb}
\usepackage{graphicx}
\usepackage{lmodern}
\usepackage{physics}
\usepackage[left=2cm,right=2cm,top=2cm,bottom=2cm]{geometry}
\usepackage{setspace}
\usepackage{booktabs}

\author{M.~O.~Musa and H.~Temimi\\
Department of Mathematics and Natural Sciences\\ Gulf University for Science and Technology\\  P.~O.~Box  7207 Hawally 32093, Kuwait}
\title{Comparison of semiclassical and quantum models of a two-level atom-cavity QED system in the strong coupling regime}
\begin{document}

\doublespacing
\maketitle

\begin{abstract}
We present a numerical study comparing  semiclassical and quantum models  of a damped, strongly interacting cavity QED system composed of a single two-level atom interacting with a single quantized cavity mode driven externally by a tunable monochromatic field.  We compute the steady state transmission spectrum of the coupled system under each model and show that in the strong coupling regime,  the two models yield starkly different results. The fully quantum mechanical model of the system correctly yields the expected multiphoton transmission spectra while the semiclassical approach results in a bistable spectrum. 
\end{abstract}

\section{Introduction}

Electromagnetic interactions between atomic transitions and quantized cavity modes alter the energy structure of both systems in a way analogous to atomic interaction in molecular formations. Cavity quantum electrodynamics, as the field is called nowadays,  has its origins in the pioneering studies of Purcell~\cite{purcell46}, Casimir and Polder~\cite{casimir48} and many other researchers~\cite{drexhage70,morawitz73,milonni73,kleppner81,gabrielse85}, leading to a flurry of experiments, including  atoms near conducting planar surfaces~\cite{hulet85,jhe87}, atoms in microwave cavities~\cite{gabrielse85,goy83} and other types of optical cavities~\cite{demartini87,heinzen87a,heinzen87b,rempe87,zhu88,raizen89,zhu90}. From the early 1990s on-wards, cavity quantum electrodynamics has rapidly grown   into one of the most vibrant sub-fields of atomic and optical physics~\cite{ berman94, haroche90,haroche06}.   

The strength of the atom-cavity interaction is often characterized using the  dimensionless cooperativity parameter  $C=g^2/2\kappa\gamma$,  where $g$ is the atom-cavity coupling strength, and $\gamma$ and $\lambda$ are the atomic and cavity decay rates, respectively. The value of the cooperativity parameter is used to demarcate the boundary between the weak  and strong coupling regimes. In the weak coupling regime the value of the cooperativity parameter is less than unity (i.e., $C\lesssim 1$\textcolor{red}{)} and dissipation  dominates coherence. In the strong coupling regime where coherence dominates dissipation and decoherence, $C\gg 1$. Early cavity QED experiments were largely confined to the weak coupling regime~\cite{kimble98} due to the low quality factor of the then available cavities. However, subsequent improvements in cavity design and laser cooling techniques shifted  the boundary of experimental research toward the  strong coupling regime, leading to observations of novel phenomena such as photon blockade~\cite{imamoglu97,birnbaum05a}, photon-pair production~\cite{kubanek08} and two-photon absorption at intensities far below levels at which nonlinear transitions normally occur~\cite{schuster08}. Recently, there has been novel trends in cavity QED research including, use of solid state photonic cavities with artificial atoms, such quantum dots in micro-pillar or micro-cavity resonators~\cite{khitrova06}; and circuit QED~\cite{blais04,wallraff04,frunzio05,koch07}, which uses a superconducting cavity coupled to charge and flux qubits, transmons, fluxoniums, quantum dots and other atom-like entities~\cite{koch07,bouchiat98,mooij99,martinis02,kitaev06,manucharyan09}. Yet another novel and promising variant of cavity QED in the optical domain which utilizes a fiber waveguide as resonator has recently been developed to generate entangled photons for quantum information processing~\cite{romero12,togan10,mirza13,su14,mirza15,masada15,aolita15,mirza15b,mirza16}. 
The novel systems offer much stronger  coupling than traditional cavity QED as well as tunability and promise a whole range of new physics, including low power nonlinear optics, become a real possibility in cavity QED systems~\cite{niemczyk10}. The field of cavity QED continues to be expanding with ever more promising systems emerging from its different variants with a push towards stronger coupling regimes where quantum mechanical effects become dominant~\cite{reithmaier04,yoshie04,peter05,khitrova06,srinivasan07,hennessy07,
ohta11,dory15,englund07, chiorescu04,forn-diaz10,yoshihara17}. In the past decade, a new subfield focused on the so-called ultrastrong coupling regime  where the atom-cavity coupling constant is comparable to a fraction of the bare atomic and cavity frequencies and the rotating wave approximation is no longer valid has emerged.~\cite{ciuti05,anappara09,devoret07,forn-diaz18,kockum18}. The emergence of new systems with with stronger coupling between two level systems and cavity modes warrants detailed modeling of the dynamics of cavity QED systems.  In this work, using a fully quantum mechanical model, we compute  the multiphoton transmission spectrum of a generic cavity QED system under moderately strong coupling regime. This work is focused on a parameter region where the cooperativity parameter is much larger than unity  (i.e., $C\gg 1$) and the coupling strength is much less than the transition and mode frequencies (i.e., $g\ll (\omega_c,\omega_a)$. We compare the results of the fully quantum mechanical model to those of the  so-called semiclassical model, showing the inadequacy of the latter model to account for the dynamics of the system beyond single photon transition driving fields.

\section{Model}
To achieve the above stated goal, we use the first state the Hamiltonian the frame co-rotating with driving field of the cavity QED system as
\begin{equation}
H=\hbar\Delta_c a^{\dagger}a+\hbar\Delta_{a}\sigma^{\dagger}\sigma+\hbar g\qty( \sigma^{\dagger}a +
\sigma a^{\dagger}).\label{JCM_hamiltonian}
\end{equation}
The first and the second terms of the Hamiltonian represent the uncoupled atom and cavity energies whereas the last term represents the interaction energy between the cavity and the atom. In addition, $\sigma=\ket{g}\bra{e}$ ($\sigma^{\dagger}=\ket{e}\bra{g}$) and $a$ ($a^{\dagger}$) are, respectively, the atomic inversion and
cavity annihilation (creation) operators. Furthermore, $\Delta_a\equiv \omega_a-\omega_l$ and $\Delta_c\equiv \omega_c-\omega_l$, where $\omega_a$ and $\omega_c$ are atomic transition and cavity mode detunings, respectively, and $\omega_a\equiv (E_e-E_g)/\hbar$ and $\omega_c$ are the atomic  and cavity resonance frequencies, and $\omega_l$ is the driving field frequency. In this work, we assume the dipole and rotating wave approximations to obtain the system Hamiltonian in Eq.~\ref{JCM_hamiltonian}. We also assume the atom  to be motionless at the antinode of the cavity mode the coupling strength $g$ is constant. Throughout this work  we assume  that values of the atom cavity coupling is much smaller than the frequencies of the atomic transition and the cavity mode. To probe its dynamical behavior, the system may be driven by coupling a tunable monochromatic field may  to the end mirror of the cavity or to the atom from the side of the cavity. In the former case an additional term representing the coupling between the cavity mode and the driving field of the form $\hbar\eta \qty(a^{\dagger}+a )$ is added to the system Hamiltonian, whereas in the second case, a term $\hbar\Omega(\sigma^{\dagger}+\sigma )$ involving the coupling between the atomic transition and the driving field is added. Here $\eta$ and $\Omega$  represent the strengths of the couplings between the driving field and the cavity or the atom. In this work we consider only the former case.
%
\subsection{Dressed states}
\begin{figure}[htbp]
\centering
\includegraphics[width=15.0cm]{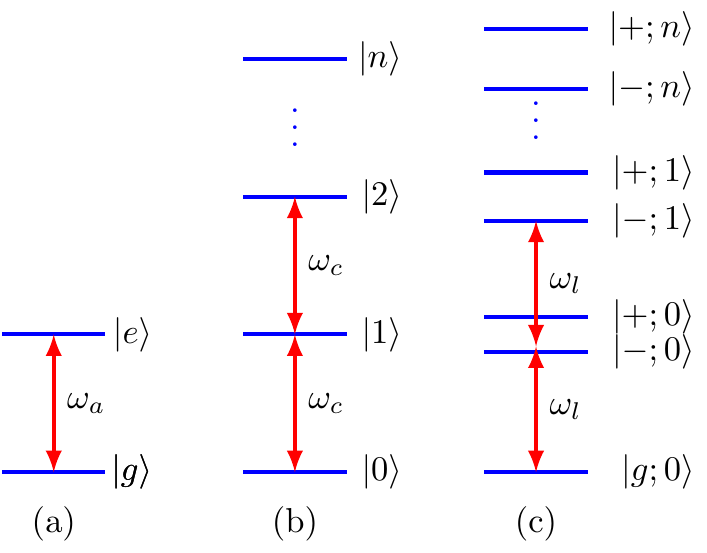}
\caption{Atom-cavity energy level structure. (a) Bare atom; (b) Bare cavity; (c) Dressed states of the atom-cavity system. Note $\omega_a=(E_e-E_g)/\hbar$, $\omega_c$ and $\omega_l$ are the atomic transition, cavity resonant and external driving frequencies. The dressed states $\ket{\pm;n}$ are linear combinations of the uncoupled states \{$\ket{g}\otimes\ket{n}$, $\ket{e}\otimes\ket{n-1}$\}, with $n=1,2, ...$ is the number of photons in the cavity.}
\label{fig:structure}
\end{figure}

\begin{figure}[htbp]
\centering
\includegraphics[width=1.0\textwidth]{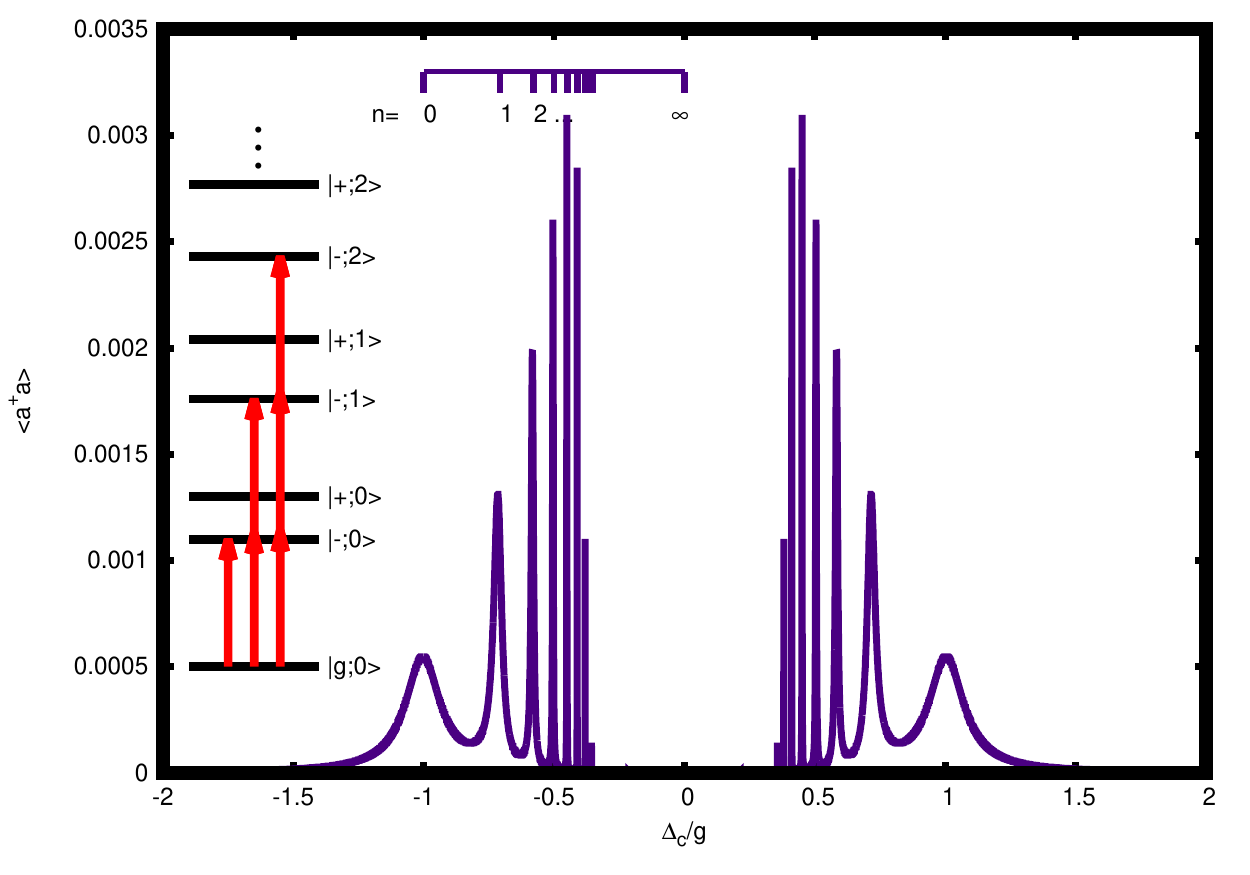}
\caption{Multiphoton transmission spectrum of strongly driven, strongly-coupled atom-cavity system. The locations of the peaks on each side $\Delta_c=0$ are given $\Delta_c=\pm g/\sqrt{n+1}$, where $n=0,1, 2, \cdots$. Energy levels and few of the lowest transitions involving one, two and three photons are indicated on the left.}
\label{fig:spec0}
\end{figure}

The energy structure of the undamped atom-cavity system in the strong coupling regime is best understood by using the dressed state approach where the atom-cavity system is treated as a single entangled system and its eigenenergies and the eigenstates are found by diagonalizing its total Hamiltonian. The diagonalization process yields
\begin{equation}
\hbar\omega_n^{\pm}=\frac{\hbar}{2}\qty[\omega_a+2\omega_c(n+\frac{1}{2})\pm\sqrt{4g^2(n+1)+\qty(\omega_a-\omega_c)^2}],
\end{equation}
 where $\hbar\omega_n^{\pm}$ are the eigenfrequencies  and  
\begin{eqnarray}
\ket{+;n}&=&\cos\theta_n\ket{e;n}+\sin\theta_n\ket{g;n+1}\\
\ket{-;n}&=&-\sin\theta_n\ket{e;n}+\cos\theta_n\ket{g;n+1},
\end{eqnarray}
are the eigenstates, where  $\ket{g;n+1}\equiv\ket{g}\otimes\ket{n+1}$ and $\ket{e;n}\equiv\ket{e}\otimes\ket{n}$ are product states of the uncoupled system (note $\ket{g}$ and $\ket{e}$ are the ground and excited states of the atom and $\ket{n}$, $n=0,1,2,...$ are the cavity Fock states, respectively). The mixing angle $\theta_n$ is defined as
\begin{equation}
\theta_n=\tan^{-1}\qty(\frac{2g\sqrt{n+1}}{(\omega_a-\omega_c)+\sqrt{4g^{2}(n+1)+\qty(\omega_a-\omega_c)^2}}).
\end{equation}
Figure~\ref{fig:structure} shows the energy structure of the coupled atom-cavity system. In the preceding analysis, we neglected the atomic and cavity dissipation which, when taken into account, render the diagonalization process intractable. Consequently,  to characterize the dynamical behavior of the system, one usually resorts to numerical methods.  Figure~\ref{fig:spec0}  illustrates numerically computed transmission spectrum of degenerately coupled (i.e., $\omega_a=\omega_c$) atom-cavity system under tunable monochromatic field.
The positions of individual peaks in each set of peaks on each side of $\Delta_c=0$, which correspond to multiphoton transitions between the ground state and pairs of dressed states, are given by the formula $\pm g/\sqrt{n+1}$ for $n=0, 1, 2, 3,\cdots$, where $n$ is the dressed state quantum number. 
\subsection{Dynamics of open atom-cavity system}
In the preceding discussion, the effect of dissipation is neglected even though it plays a central role in shaping the behavior of the coupled atom-cavity system. Any realistic analysis of the dynamical behavior of such system has to take dissipation into account and the most direct way to achieving this is to use density operator formalism. The density operator of a system obeys the master equation~\cite{carmichael99}
\begin{eqnarray}
\dot{\rho}=&-&\frac{i}{\hbar}\comm{H}{\rho}+\frac{1}{2}\kappa\qty(2a\rho a^{\dagger}-a^{\dagger}a\rho-\rho a^{\dagger}a)
+\frac{1}{2}\gamma\qty(2\sigma\rho\sigma^{\dagger}-\sigma^{\dagger}\sigma\rho-\rho\sigma^{\dagger}\sigma).
\label{eq:master_equation}
\end{eqnarray}
In the above master equation $H$ is the total Hamiltonian of the undamped system whereas the second and third terms represent the cavity and the atom dissipation, respectively, and the parameters  $\gamma$ and $\kappa$ are the damped rates defined previously. The density operator offers a means of calculate statistical averages of the system operators. With knowledge of the density operator, For example, mean value of system operators as well as their products can be calculated. For example, the mean value of operator $\mathcal{O}$ is obtained by calculating the expectation value of the product of $\mathcal{O}$ and the density operator. In addition, the time evolution of the expectation value of $\mathcal{O}$ can be obtained by using Equation~\ref{eq:master_equation} and the cyclic properties of the trace such that $\ev{\dot{\mathcal{O}}}=\Tr (\mathcal{O}\dot{\rho})$, where $\Tr$ stands for trace operation and $\dot{\rho}$ obeys the master equation~\cite{carmichael99}. In the case of the steady state properties of the system, the steady state value of the density operator is used to calculate steady state expectation values of system operators.
\subsection{Equations of motion}
The master equation provides a means for deriving dynamical equations for the expectation values of any of the the system operators. The complete dynamics of the atom-cavity system are contained in the equations of motion of the expectation values of the atomic inversion $\sigma$, population inversion $\sigma_z$ and the cavity field $a$ operators. These operators obey the set of nonlinear, coupled differential equations
\begin{eqnarray}
\ev{\dot{\sigma}}&=&-i(\Delta_a-\frac{i}{2}\gamma)\ev{\sigma}+ig\ev{a\sigma_z}\label{eq:coupled_eqs1}  \\
\ev{\dot{\sigma}_z}&=&-\gamma\qty(\ev{\sigma_z}+1)+2ig\qty(\ev{\sigma a^{\dagger}}-\ev{a\sigma^{\dagger}})\label{eq:coupled_eqs2} \\ 
\ev{\dot{a}}&=&-i\qty(\Delta_c-\frac{i}{2}\kappa)\ev{a}-ig\ev{\sigma}-i\eta.
\label{eq:coupled_eqs3}
\label{eq:coupled_eqs}
\end{eqnarray}
Under the so-called semiclassical approximation, the above equations are solved by approximating the expectations of the products of the cavity and atomic operators  with products of their expectations such as $\ev{a\sigma_z}\approx\ev{a}\ev{\sigma_z}$. The semiclassical approximation is a valid under conditions where multiphoton excitation are not possible either because of  dissipation dominating coherences or weak external driving. Otherwise, semiclassical model breaks down and a fully quantum mechanical analysis is necessary. Under the fully quantum mechanical model,  expectations of products such as $\ev{a\sigma^{\dagger}}$ need to be treated as as new operators which satisfy their own equations of motion. For example, $\ev{a\sigma^{\dagger}}$ obeys the equation of motion
\begin{eqnarray}
\dv{t}\ev{a\sigma^{\dagger}}&=&-i\qty(\Delta_c+\Delta_a-i(\gamma-\frac{1}{2}\kappa))\ev{\sigma^{\dagger}a}
-ig\ev{\sigma_z a^{\dagger}a}-\frac{i}{2}g(\ev{\sigma_z}+1)-i\eta\label{eq:coupled_eqs5}
\end{eqnarray}
which involves the higher order operator product $\ev{\sigma_z a^{\dagger}a}$, which in turn obeys the differential equation
\begin{eqnarray}
\dv{t}\ev{\sigma_z a^{\dagger}a}&=&-(\kappa+\gamma)\ev{\sigma_z a^{\dagger}a}-\gamma\ev{a^{\dagger}a}-2ig\ev{\sigma (a^{\dagger})^2 a} 
+ig\ev{\sigma a^{\dagger}}-i\eta(\ev{\sigma_z a^{\dagger}}-\ev{\sigma_z a}).\label{eq:coupled_eqs6}
\end{eqnarray}
This equation of motion too involves the new operator products $\ev{a^{\dagger}a}$  and $\ev{\sigma (a^{\dagger})^2 a}$ whose equations of motion involve operator products involving higher powers of $a$ and $a^{\dagger}$ and the cavity field operators. Accordingly, the number of equations of motion to fully characterize the behavior of the system is potentially infinite. Under sufficiently weak driving conditions, only the lowest energy transitions occur and the process of producing new operator products terminates due to the fact that $(a^{\dagger})^{m}a^{n}\ket{i;n^{\prime}}=0$ for some $n>n^{\prime}$. However, in general, the  solution of the system involves potentially infinite number of coupled differential equations.  In this work, we calculate the steady-state transmission spectrum of the system using the both the semiclassical and fully quantum mechanical models. In the quantum mechanical model, we calculate numerically the steady state expectation values of the atomic and cavity operators as functions of the cavity detuning $\Delta_c$ under various pump and dissipation conditions. In order to achieve this goal, we 
first calculate the steady state density matrix of the system iteratively using the master equation (Eq.~\ref{eq:master_equation}) by utilizing the Quantum Toolbox in Python (QuTiP) framework~\cite{johansson13}. We then calculate the expectation values of the system operators using the steady state density matrix. We also calculate the semiclassical steady-state transmission of the system as function of the detuning $\Delta_c$ by numerically solving factorized version of Equations~\ref{eq:coupled_eqs1} -~\ref{eq:coupled_eqs3} for the expectation of the cavity field operator $a$ with all the derivatives set to zero.
\section{Results}
\begin{figure}
\includegraphics[width=1.0\textwidth]{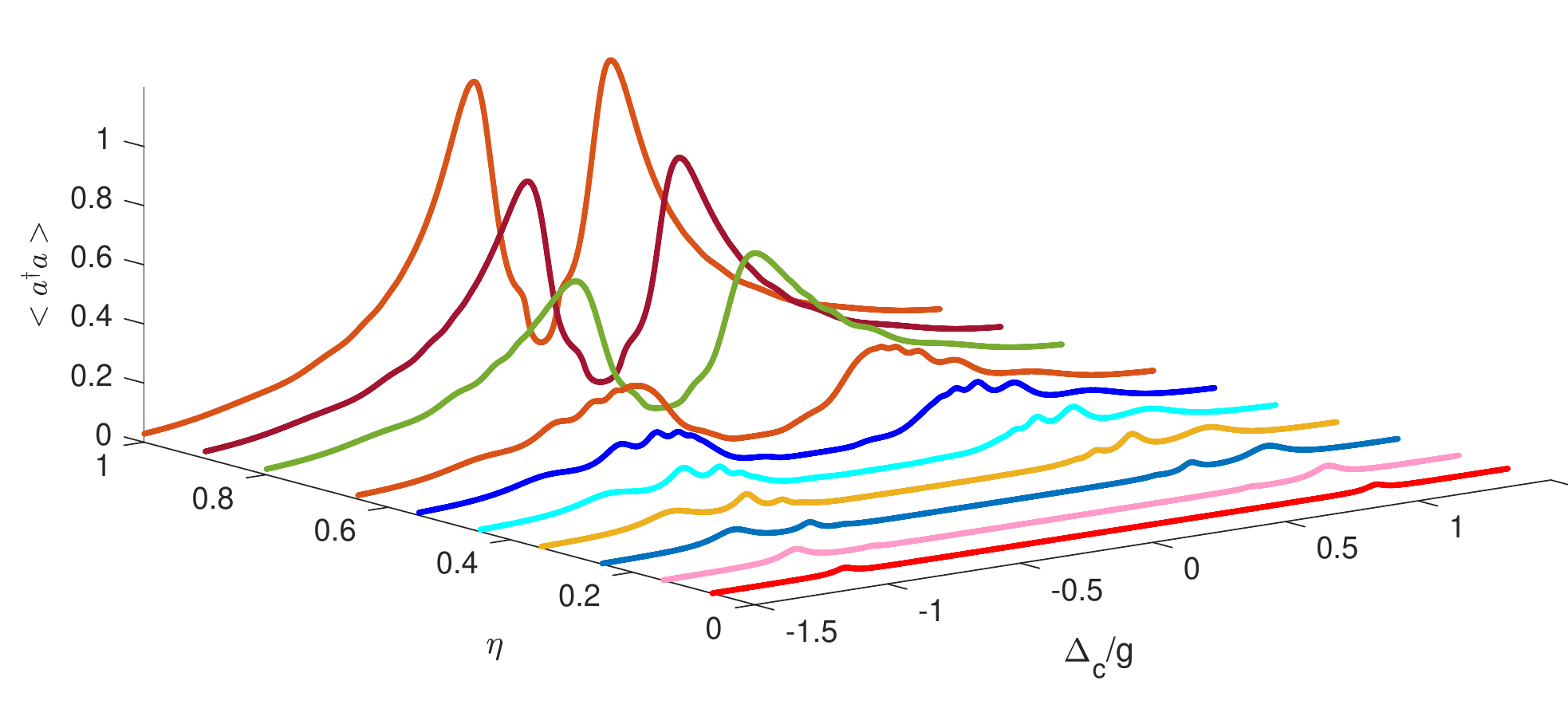}
\caption{Steady state transmission intensity ($I\propto \ev{a^{\dagger}a}$) of the quantum mechanical model as function of driving field detuning ($\Delta_c$) 
for parameters $\kappa=\gamma=g/3=1$ (i.e.,  $C=g^2/2\gamma\kappa=4.5$). Different curves in the Figure correspond to different values of the driving field strength $\eta$ between 0.1 and 1.0. The pair of peaks in each curve correspond to transitions between the ground state and the first pair of dressed states of the coupled system. }
\label{fig:spec1}
\end{figure}

\begin{figure}
\includegraphics[width=1.0\textwidth]{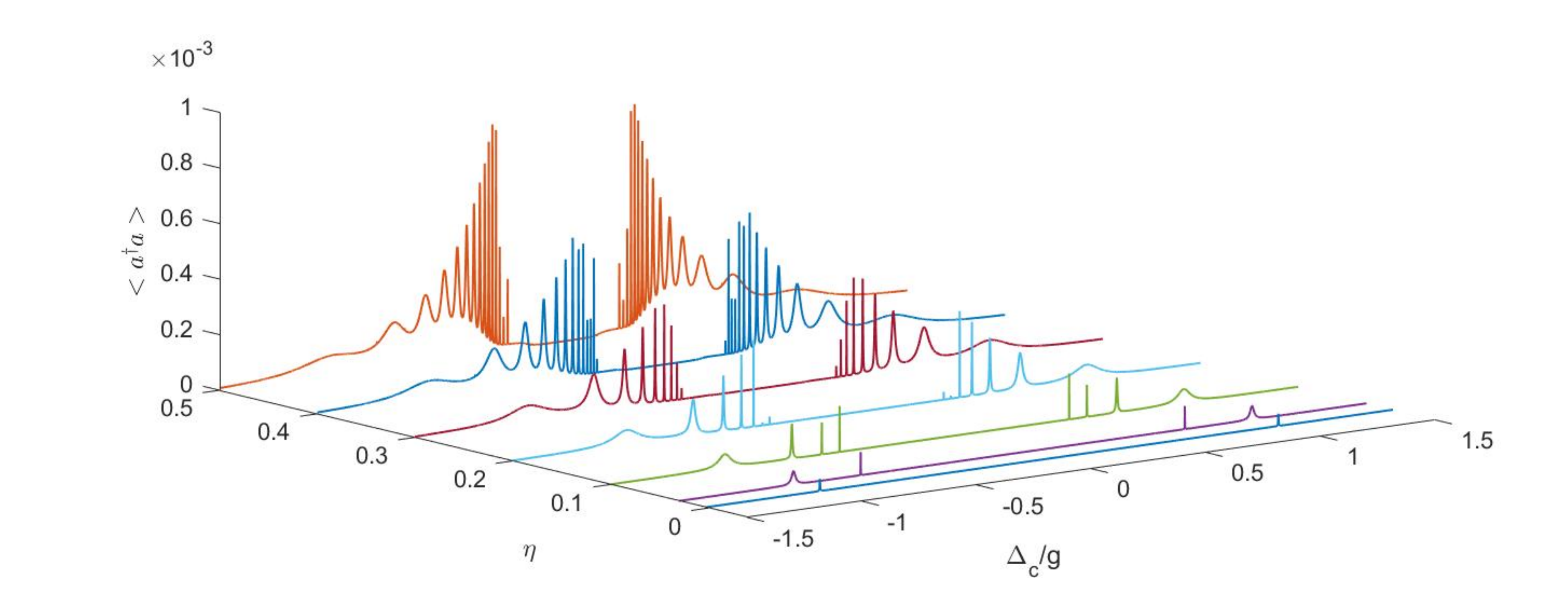}
\caption{Steady state transmission intensity ($I\propto \ev{a^{\dagger}a}$) of the coupled atom-cavity system based on the quantum mechanical model as function of driving field detuning ($\Delta_c$) 
for parameters $\kappa=\gamma=0.0083g$ (i.e.,  $C=g^2/2\gamma\kappa=4.5\times 10^6$). Different curves correspond to different values of the driving field strength $\eta$. As shown, two new peaks appear gradually at $\Delta_c=g/\sqrt{2}$ and $\Delta_c=g/\sqrt{3}$ as the relative value of driving increases from 0.005 to 0.5}
\label{fig:spec2}
\end{figure}

\begin{figure}
\includegraphics[width=1.0\textwidth]{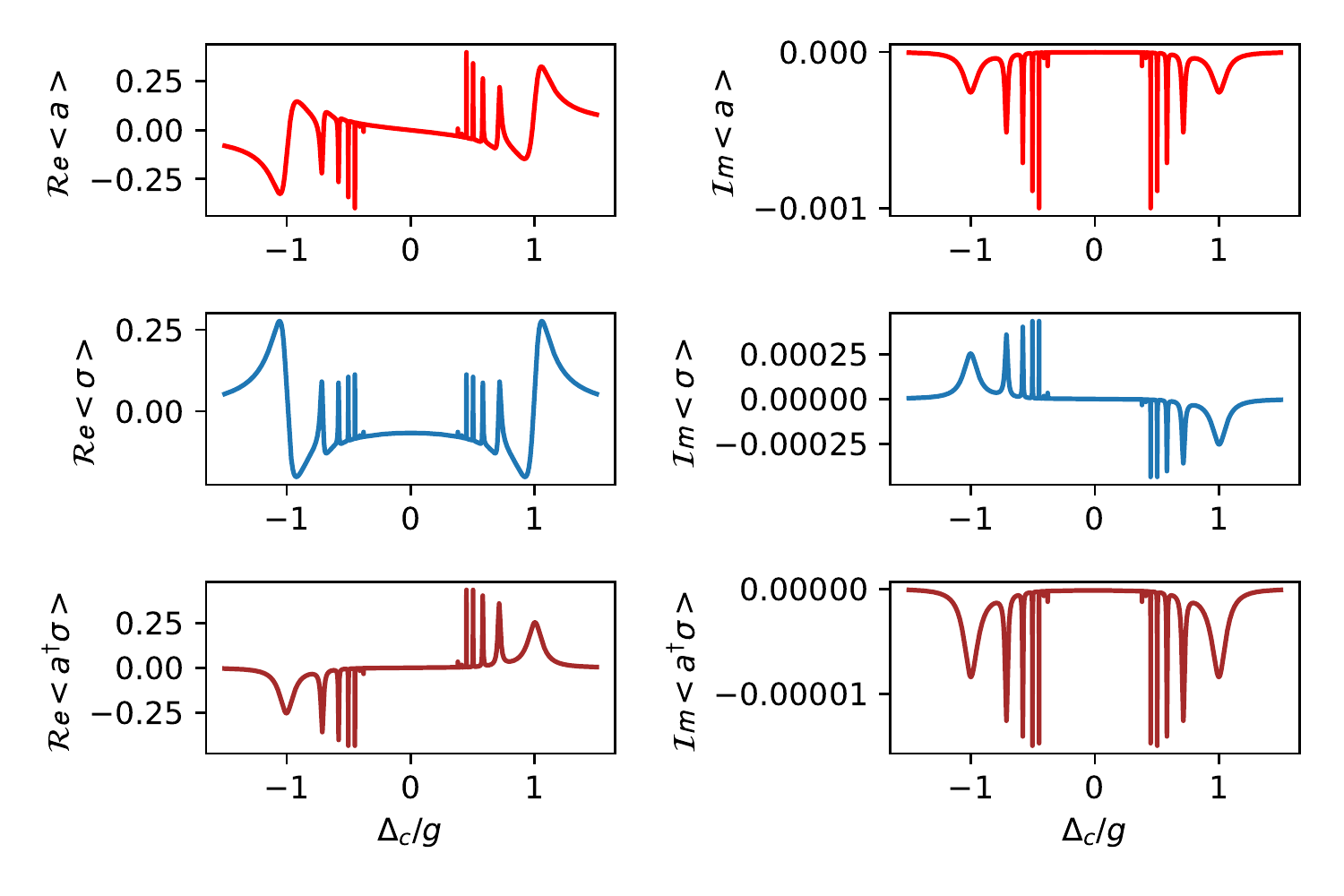}
\caption{Real and imaginary parts of $\ev{a}$, $\ev{\sigma}$ and $\ev{a^{\dagger}\sigma}$ in the strong coupling regime under strong driving conditions.}
\label{fig:spec3}
\end{figure}

Figures~\ref{fig:spec1} and~\ref{fig:spec2} show computed quantum mechanical transmission spectra of the resonantly-coupled atom-cavity system as function of the cavity detuning ($\Delta_c$) for different levels of dissipation and external driving field conditions. The computed intensity transmission of the cavity is proportional to the steady-state mean photon number $\ev{a^{\dagger}a}$ in the cavity.  The individual curves in each of figure represent the cavity  transmission under same atomic and cavity parameters but at different levels of driving field strength. The spectra in Figure~\ref{fig:spec1} are based on atomic and cavity parameters of $\kappa/g=\gamma/g=1/3$ (corresponding to cooperative parameter value of $C=4.5$) with the different curves corresponding to relative driving field strength (i.e., $\eta/g$) values ranging from $0.05$ to $1.0$. On the other hand, the spectra in  Figures~\ref{fig:spec2} correspond to much lower atomic and cavity dissipation rates (stronger coupling) amounting to a cooperative parameter of $C=4.5\times 10^{6}$ with the relative driving field strength  $\eta/g$ ranging from $0.05$ to $0.5$. Figure~\ref{fig:spec3} shows the plots of the real and imaginary parts of the cavity and atomic operators $\ev{a}$, $\ev{\sigma}$ and $\ev{a^{\dagger}\sigma}$ as functions of the driving field detuning. The data in the figure is computed using the fully quantum mechanical model in the strong coupling regime.

\begin{figure}
\includegraphics[width=1.0\textwidth]{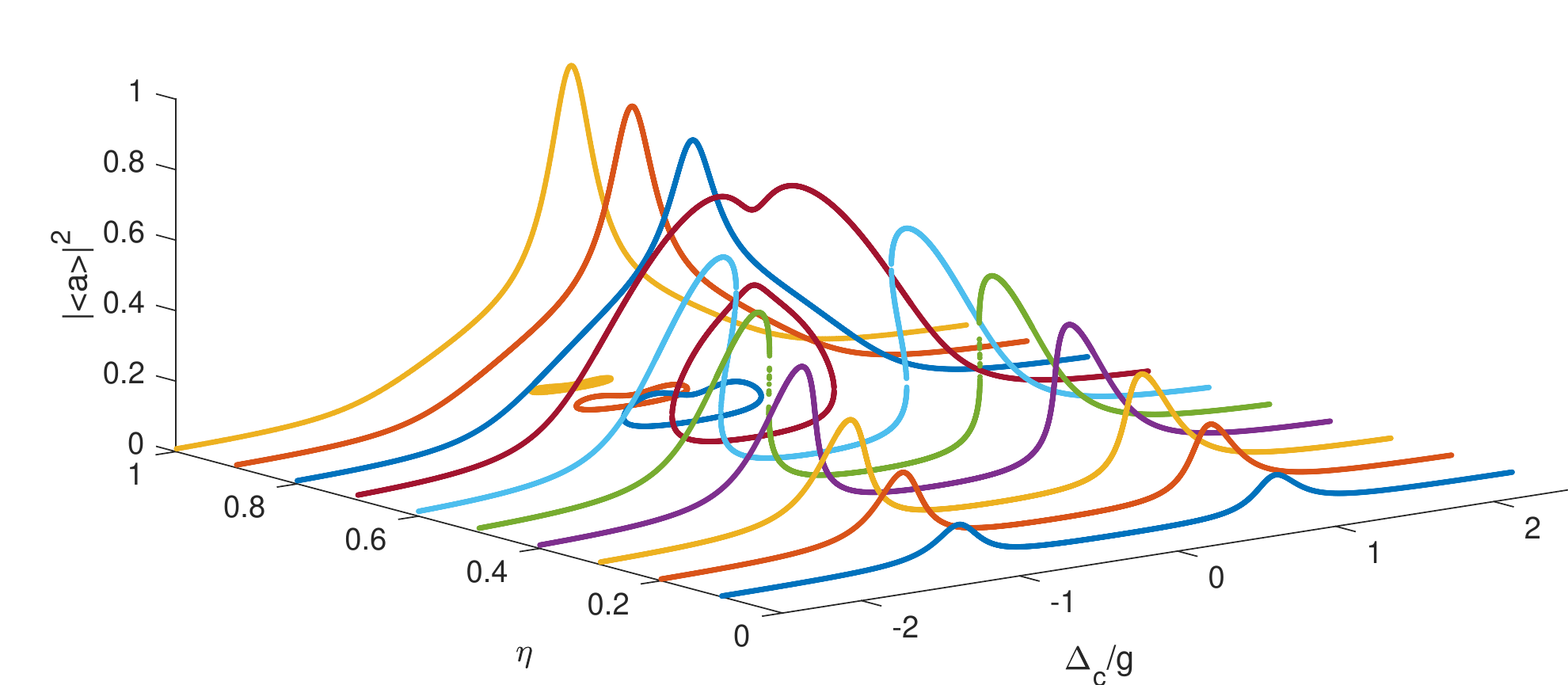}
\caption{The curves  correspond to the calculated transmission intensity of the cavity based on the semiclassical model for system parameter values of $\gamma/g=\kappa/g=1/3$ and $\eta/g$ ranging from 0.03 to 0.3.}
\label{fig:spec4}
\end{figure}

\begin{figure}
\includegraphics[width=1.0\textwidth]{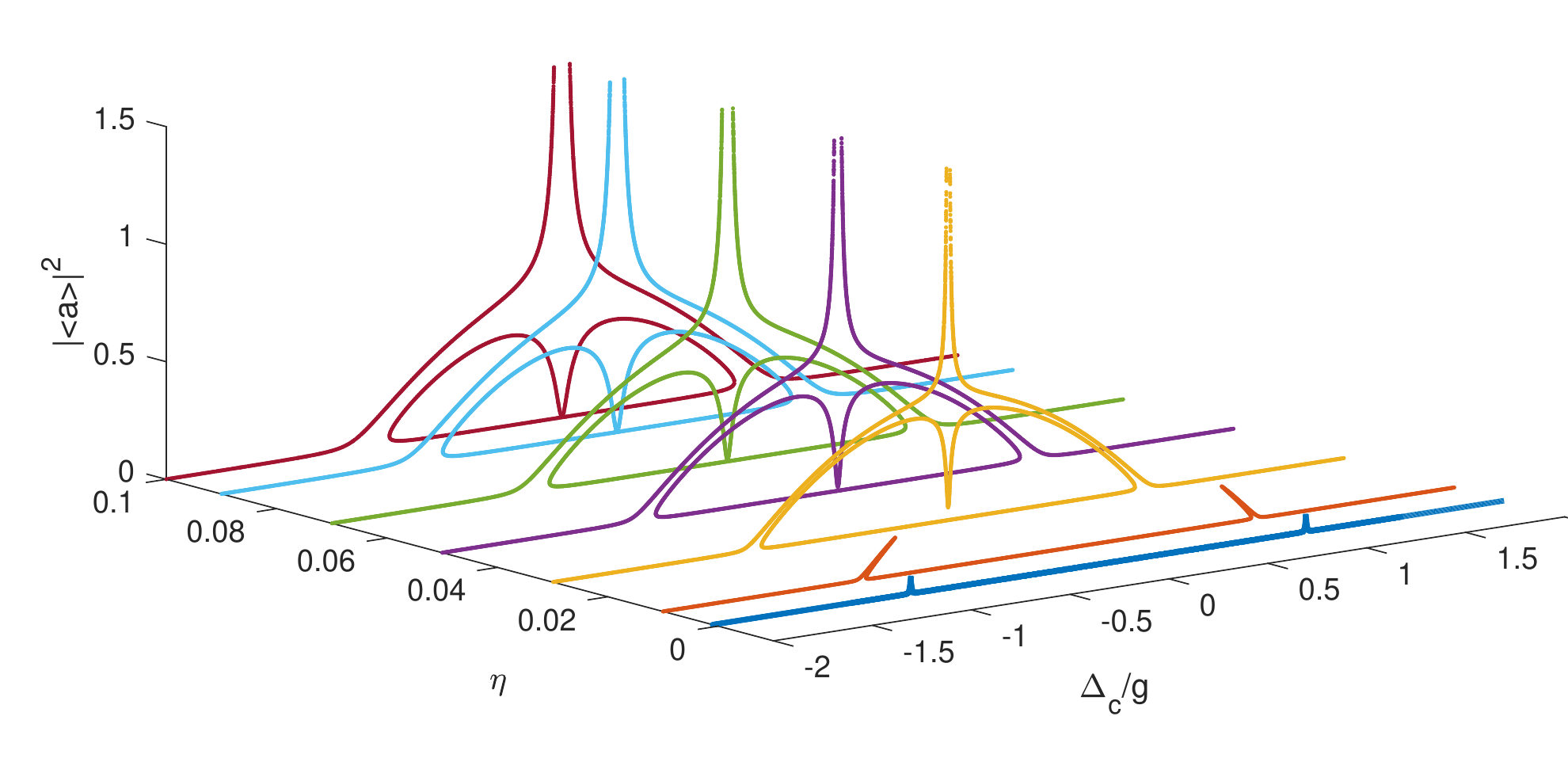}
\caption{The curves correspond to the calculated transmission intensity of the cavity based on the semiclassical model for system parameter values of $\kappa/g=\lambda/g=0.0083$ and relative driving field strength $\eta/g$ ranging from $3.3\times 10^{-4}$ to $6.7\times 10^{-2}$}
\label{fig:spec5}
\end{figure}

\begin{figure}[htbp]
\centering
\includegraphics[width=1.0\textwidth]{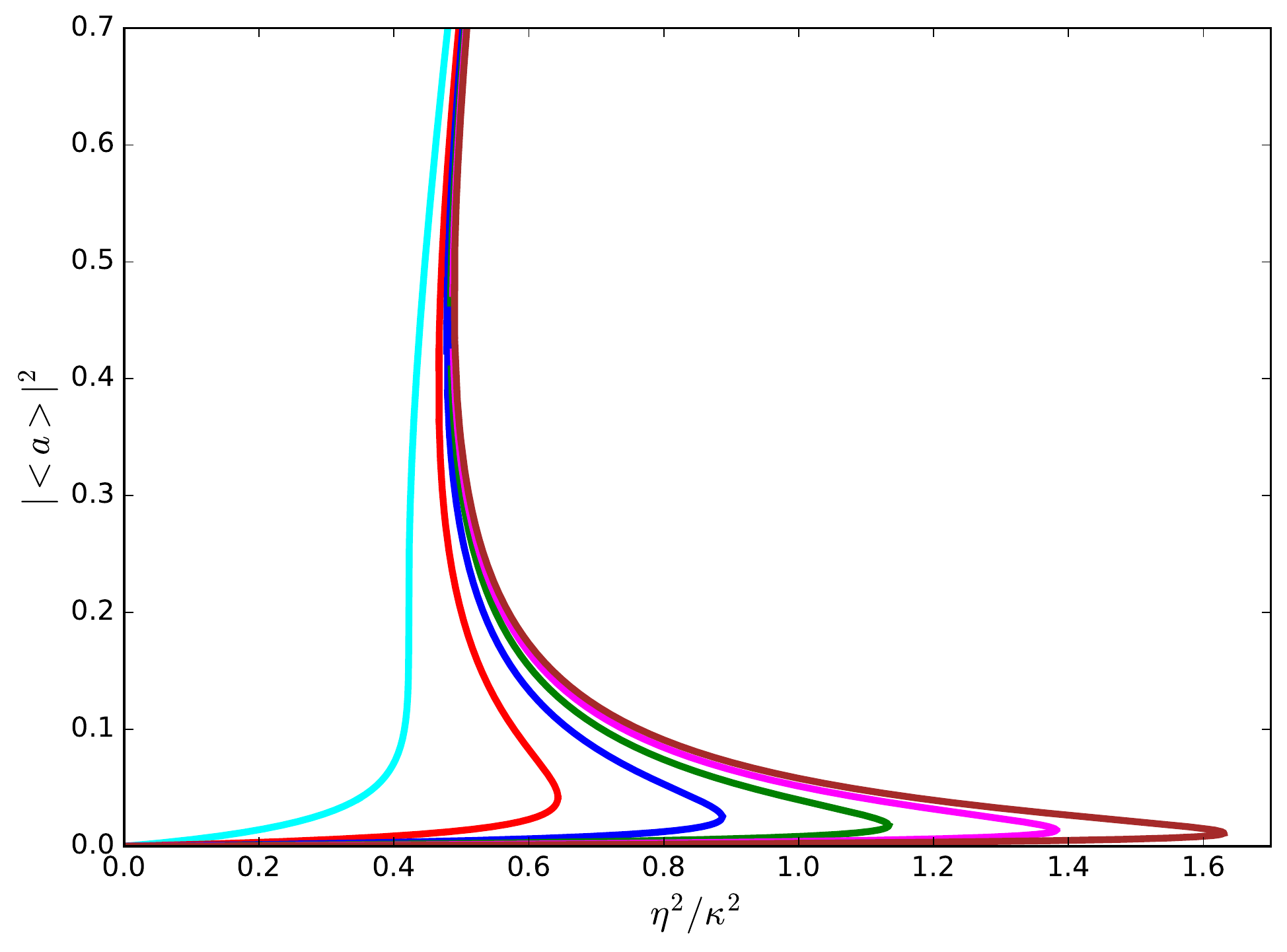}
\caption{Bistability curves for various values of $C=g^/2\lambda\gamma$ under the factorization assumption.}
\label{fig:bistability}
\end{figure}

Furthermore, Figures~\ref{fig:spec4} and~\ref{fig:spec5} show calculated intensity transmission of the cavity based on the semiclassical model. The atomic and cavity dissipation parameters  used in the calculations are chosen to be similar to those used to calculate the spectra in Figures~~\ref{fig:spec1} and~\ref{fig:spec2}.  The curves in Figure~\ref{fig:spec4} correspond to system parameter values of $\gamma/g=\kappa/g=1/3$ and $\eta/g$ ranging from 0.03 to 0.3 whereas the data in Figure~\ref{fig:spec5} correspond to system parameters of $\kappa/g=\lambda/g=0.0083$ and relative driving field strength $\eta/g$ ranging from $3.3\times 10^{-4}$ to $6.7\times 10^{-2}$. The semiclassical spectra in these figure are multivalued functions of the detuning. Finally, Figure~\ref{fig:bistability} shows the calculated transmission of the cavity as function of the driving field strength. The different curves in the figure  correspond to cooperativity parameter values ranging from $1$ to $6$. The semiclassical transmission spectra of the cavity show bistable at higher driving fields  at sufficiently low dissipation rates.  


\section{Discussion}

Figures~\ref{fig:spec1},~\ref{fig:spec2},~\ref{fig:spec4} and~\ref{fig:spec5} present transmission of spectra of the quantum mechanical and semiclassical models of the resonantly-coupled atom-cavity system under different levels of driving field strength and dissipation conditions. The fully quantum mechanical  spectra, shown in Figures~\ref{fig:spec1} and \ref{fig:spec2},  are based on computation of the steady-state mean cavity photon number $\ev{a^{\dagger}a}$, whereas the semiclassical spectra in Figures~\ref{fig:spec4} and~\ref{fig:spec5} are based on the squared modulus of the  steady-state cavity field amplitude (i.e., $|\ev{a}|^2$). As shown in the figures, under sufficiently weak driving (small $\eta$), irrespective of the coupling strength and damping conditions, the semiclassical and the fully quantum mechanical models yield indistinguishable, double-peaked transmission spectra.  However, under stronger driving field conditions, the two models yield starkly different results, with the quantum mechanical model resulting in a multi-peaked (multiphoton) transmission spectrum and the semiclassical model analysis resulting in a bistable spectrum. 

In the fully quantum mechanical case, the two sets of peaks  on each side of $\Delta_c=0$ in the spectra in Figures~\ref{fig:spec1} and~\ref{fig:spec2} correspond to transitions between the ground state $\ket{g;0}$  and pairs of dressed states $\ket{\pm;n}$ of the coupled system, where $n=0,~1,~2,~\hdots$  The peak positions in the spectra are determined based on the  resonance condition $(n+1)\omega_l=(n+1)\hbar\omega_c\pm g\sqrt{n+1}$. The transition energies of  $n$-photon resonances are shifted down by  subtraction of $n\hbar\omega_l$ in order to present the spectra in the  compact (folded) form of Figures~\ref{fig:spec1} and~\ref{fig:spec2}. The locations of individual peaks in the spectra are determined according to the condition $\Delta_c=\pm g/\sqrt{n+1}$ which implies that the spacing of the peaks in the spectra decreases as $1/\sqrt{n+1}$.  In general, the overall resolution of the spectrum depends on both the strength of the atom-cavity coupling,  dissipation rates as well as the driving field strength. The spacing of the two branches of the spectrum and also the peaks within each branch depends primarily on the strength of the coupling ($g$) between the cavity mode and the atomic transition. The stronger the coupling,  the larger the spacing between the two branches of the spectrum, and also the spacing between the  peaks within each branch.  For a fixed value of the coupling constant, the widths of the individual transitions and, consequently the overall resolution of the spectrum, depends on the value of the dissipation rates ($\kappa$, $\gamma$).  For the case where dissipation dominating coherence, the multiphoton transition peaks are not resolved, as shown in Figure~\ref{fig:spec1}. High dissipation degrades the resolution as well as the efficiency of multiphoton transitions. On the other hand, when coherence dominates the dissipation, the efficiency for the system to absorp multiple photons simultaneously becomes high as indicated by the multiphoton spectra in~Figure~\ref{fig:spec2}. In addition to exciting multiphoton transitions, strong driving field also saturates lower transitions. Table~\ref{tab:peak_width} examines the saturating effect of the driving field strength on the width of the lowest (single) photon transitions. Using the data in the Table we infer the onset of the broadening for single photon transitions to correspond to  $\eta\sim 2.0\times 10^{-4}g$, which corresponds to an average cavity photon number of $\ev{a^{\dagger}a}\sim 4.3\times 10^{-7}$. This result is very close to the theoretical single photon transition saturation mean photon number of $n_0=4\gamma^2/g^2\sim 3.3\times 10^{-7}$~\cite{kimble98}.  Beyond this intensity, the transitions does not absorb any more radiation and the incident driving field gets completely reflected by the cavity.  This effect, termed as photon blockade, was first reported by Imamoglu \textit{et al.} in 1997~\cite{imamoglu97} and has been observed in other  cavity QED experiments (e.g., Ref.~\cite{birnbaum05a}).
\begin{table}[h]
\centering
\resizebox{\textwidth}{!}{
\begin{tabular}{ccc|ccc}
\toprule
$\eta $ & Peak Area & Peak Width&$\eta$ & Peak Area & Peak Width\\
\midrule
$0.0001$ & $1.00\times 10^{-9}$& $5.001\times 10^{-3}$&$0.0040$ & $1.57\times 10^{-6}$& $6.403\times 10^{-3}$\\
$0.0002$ & $4.00\times 10^{-9}$& $5.004\times 10^{-3}$&$0.0080$ & $5.18\times 10^{-6}$& $9.434\times 10^{-3}$\\
$0.0004$ & $1.80\times 10^{-8}$ & $5.016\times 10^{-3}$ & $0.0200$ & $2.19\times 10^{-5}$& $2.062\times 10^{-2}$\\
$0.0008$ & $7.10\times 10^{-8}$ & $5.064\times 10^{-3}$ & $0.0400$ & $6.26\times 10^{-5}$ & $4.031\times 10^{-2}$ \\
$0.0020$ & $4.28\times 10^{-7}$ & $5.385\times 10^{-3}$ & $0.0800$ & $1.78\times 10^{-4}$& $8.017\times 10^{-2}$\\
\bottomrule
\end{tabular}
}
\caption{Variation of the area under peaks corresponding to single photon transitions between the lowest pair of dressed states $\ket{\pm;0}$ and the ground state as function of the driving field strength.}
\label{tab:peak_width}
\end{table}

Figures~\ref{fig:spec4} and~\ref{fig:spec5} show the  semiclassical steady-state transmission spectrum of the atom-cavity system at various levels of the driving field strength. The spectra in Figure~\ref{fig:spec4} correspond to a cooperativity parameter in the borderline between weak and strong coupling regimes ($C=4.5$),  whereas the data in Figure~\ref{fig:spec5} correspond to the system parameters deeply in the strong coupling regime ($C=7.26\times 10^3$). Under low driving conditions, both spectra consist a pair of peaks that get bent inward as the driving field strength increases and, unlike the fully quantum mechanical case, there are no multiphoton peaks. As the driving field increases further, the pair of peaks get distorted futher and develop a lope around $\Delta_c=0$ before joining into a single, very wide peak centered at $\Delta_c=0$.
The multivalued nature of the semiclassical spectrum under high driving conditions is evidence of a single atom bistability. Figure~\ref{fig:bistability} makes this point more evident by plotting the cavity transmission as function of the driving field intensity. These bistability curves, which  are computed  by  solving the the steady state, factorized form of Eqs.~\ref{eq:coupled_eqs1}-~\ref{eq:coupled_eqs3} for $\ev{a}$, are single-atom analog of the bistability curves for atomic ensembles in cavities~\cite{lugiato84}. Comparing the semiclassical and quantum mechanical calculations, it is evident that they lead to quite different results under sufficiently strong driving field conditions. The quantum model leads to multiphoton spectrum whereas the semiclassical model predicts the cavity transmission to be bistable. The semiclassical model neglects the coherences between the atomic and cavity operators. As a result, it yields good results only if the system is in the weak coupling regime where dissipates coherence or if the  driving field is sufficiently weak such that multiphoton transitions are not excited. When these conditions do not hold, the semiclassical model does not predict the spectrum of the system correctly. 

Lastly, it is worth commenting on the behavior of the atomic-cavity field operators, namely, $\ev{a}$, $\ev{\sigma}$ and $\ev{a^{\dagger}\sigma}$. Figure~\ref{fig:spec3} shows the dependence of these operators on the the detuning and two interesting points may be notes. First, the real and imaginary parts of the mean value of the cavity field annihilation operator shows normal behavior. The real part of $\ev{a}$ shows dispersive character whereas the imaginary part shows absorptive character.  On the other hand, the real part of $\ev{\sigma}$ show dispersive behavior without the $\pi$ phase change across $\Delta_c$ whereas the imaginary part shows absorptive behavior with a phase change of $\pi$ across $\Delta_c$. Lastly, both the real and imaginary parts of $\ev{a^{\dagger}\sigma}$ show absorptive behavior. In all cases, the imaginary part has a much less value than the real part.
\section{Conclusions}
In this work, we explored numerically the behavior of generic single-atom cavity QED system using both semiclassical and fully quantum mechanical models and to show that the two approaches yield similar results under sufficiently low driving fields. More importanly, we showed that the two models yield starkly different results when the system is  deeply in the  strong coupling regime. Our analysis shows the semiclassical model to be unsuitable under such condition.   Whereas the fully quantum mechanical model correctly predicts in a multiphoton spectra when the system in the deeply in the strong coupling regime, the semiclassical model incorrectly predicts single atom bistable spectra. Therefore, under conditions where the system parameters are sufficiently in the strong coupling regime, fully quantum mechanical analysis is warranted.

\end{document}